\documentclass[a4paper]{jpconf}
\usepackage{graphicx}
\usepackage{color}
\begin{document}
\title{LBECA: A Low Background Electron Counting Apparatus for Sub-GeV Dark Matter Detection}

\author{A. Bernstein$^a$, M. Clark$^b$, R. Essig$^c$, M. Fernandez-Serra$^d$, A. Kopec$^b$, R.F. Lang$^b$, J. Long$^e$\footnote{Now at the University of Chicago}, K. Ni$^{e*}$, S. Pereverzev$^a$, J. Qi$^e$, P. Sorensen$^f$, Y. Wei$^e$, J. Xu$^a$, J. Ye$^e$ and C. Zhen$^{c,d}$}

\address{$^a$Lawrence Livermore National Laboratory (LLNL), Livermore, CA 94550-9698, USA}
\address{$^b$Department of Physics and Astronomy, Purdue University, West Lafayette, IN 47907, USA}
\address{$^c$C.N. Yang Institute for Theoretical Physics, Stony Brook University, Stony Brook, NY 11794, USA}
\address{$^d$Department of Physics and Astronomy, Stony Brook University, Stony Brook, NY 11794, USA}
\address{$^e$Department of Physics, University of California San Diego, La Jolla, CA 92093, USA}
\address{$^f$Lawrence Berkeley National Laboratory (LBNL), Berkeley, CA 94720-8099, USA}
\ead{nikx@physics.ucsd.edu}

\begin{abstract}
Two-phase noble liquid detectors, with large target masses and effective background reduction, are currently leading the dark matter direct detection for WIMP masses above a few GeV. Due to their sensitivity to single ionized electron signals, these detectors were shown to also have strong constraints for sub-GeV dark matter via their scattering on electrons. In fact, the most stringent direct detection constraints for sub-GeV dark matter down to as low as ~5 MeV come from noble liquid detectors, namely XENON10, DarkSide-50, XENON100 and XENON1T, although these experiments still suffer from high background at single or a few electron level. LBECA is a planned 100-kg scale liquid xenon detector with significant reduction of the single and a few electron background. The experiment will improve the sensitivity to sub-GeV dark matter by three orders of magnitude compared to the current best constraints.
\end{abstract}

\section{Introduction}
The dark matter in the mass range above a few GeV is currently heavily probed by experiments using liquid xenon. The upcoming 10-tonne scale liquid xenon detectors, PandaX-4T, XENONnT and LZ and eventually the Generation-3 detector, liekly the proposed DARWIN experiment, will be capable of probing Weakly Interacting Massive Particles (WIMPs) down to the signal from atmospheric neutrinos. With the WIMP model coming under such strong pressure, the community has been working in parallel on detector concepts that can probe dark matter over a much wider mass range, especially several well-motivated dark matter candidates below the proton mass~\cite{Battaglieri:2017aum,BRNreport}.

The traditional direct detection technique---observing nuclear recoils from dark matter scattering elastically off nuclei---rapidly loses sensitivity in existing experiments for dark matter masses below $\sim$1~GeV. However, a technique with significant potential for improvement is to search for dark matter scattering off electrons~\cite{Essig:2011nj} using noble liquid detectors, as first demonstrated using data from XENON10~\cite{Essig:2012yx,Angle:2011th}, XENON100~\cite{Essig:2017kqs,Aprile:2016wwo} and DarkSide-50~\cite{Agnes:2018oej} experiments. Solid-state silicon detectors, such as SENSEI~\cite{Crisler:2018gci,sensei2019}, DAMIC~\cite{Aguilar-Arevalo:2019wdi}, and  SuperCDMS~\cite{Agnese:2018col} have also demonstrated their potential in the region for sub-GeV light dark matter detection. More recently, constraints on sub-GeV dark matter from electronic recoils were further improved by the XENON1T~\cite{Aprile:2019xxb} experiment. The main factors currently limiting the sensitivity of liquid xenon detectors are previously-ignored background sources at the level of single or a few electrons. By mitigating and controlling these backgrounds, liquid xenon detectors will play a crucial role in probing sub-GeV dark matter.  

\section{The Signal and Background of a Few Electrons}
Two-phase LXe detectors extract two signals for each particle interaction in their target: One is the prompt scintillation signal (S1). The other is a delayed signal from ionization (S2): electrons liberated in the interaction are drifted up to the liquid/gas interface, where a strong electric field extracts them into the gas phase and causes a proportional scintillation signal. The energy threshold of the two-phase LXe detectors in standard analyses is set by the S1 light collection efficiency to 2-3 photoelectrons (PE). In contrast, thanks to the inherent S2 amplification, even a single extracted electron can produce $\sim$20-100~PE in S2, which thus allows for a much lower energy threshold when using this channel alone (``S2-only" analysis).

\begin{figure}[!htb]
\begin{center}
    \includegraphics[width=0.45\textwidth]{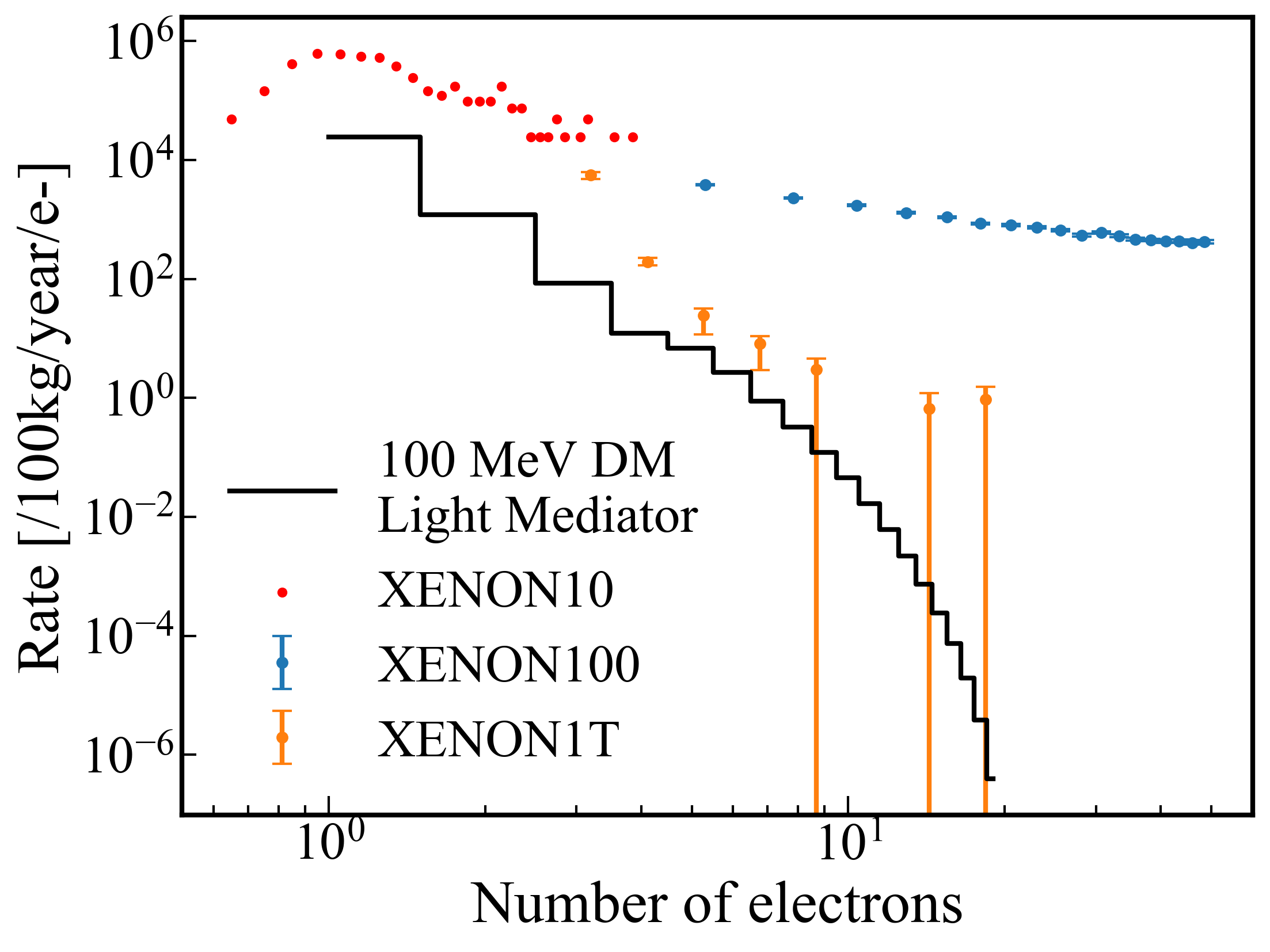}
    \includegraphics[width=0.4\textwidth]{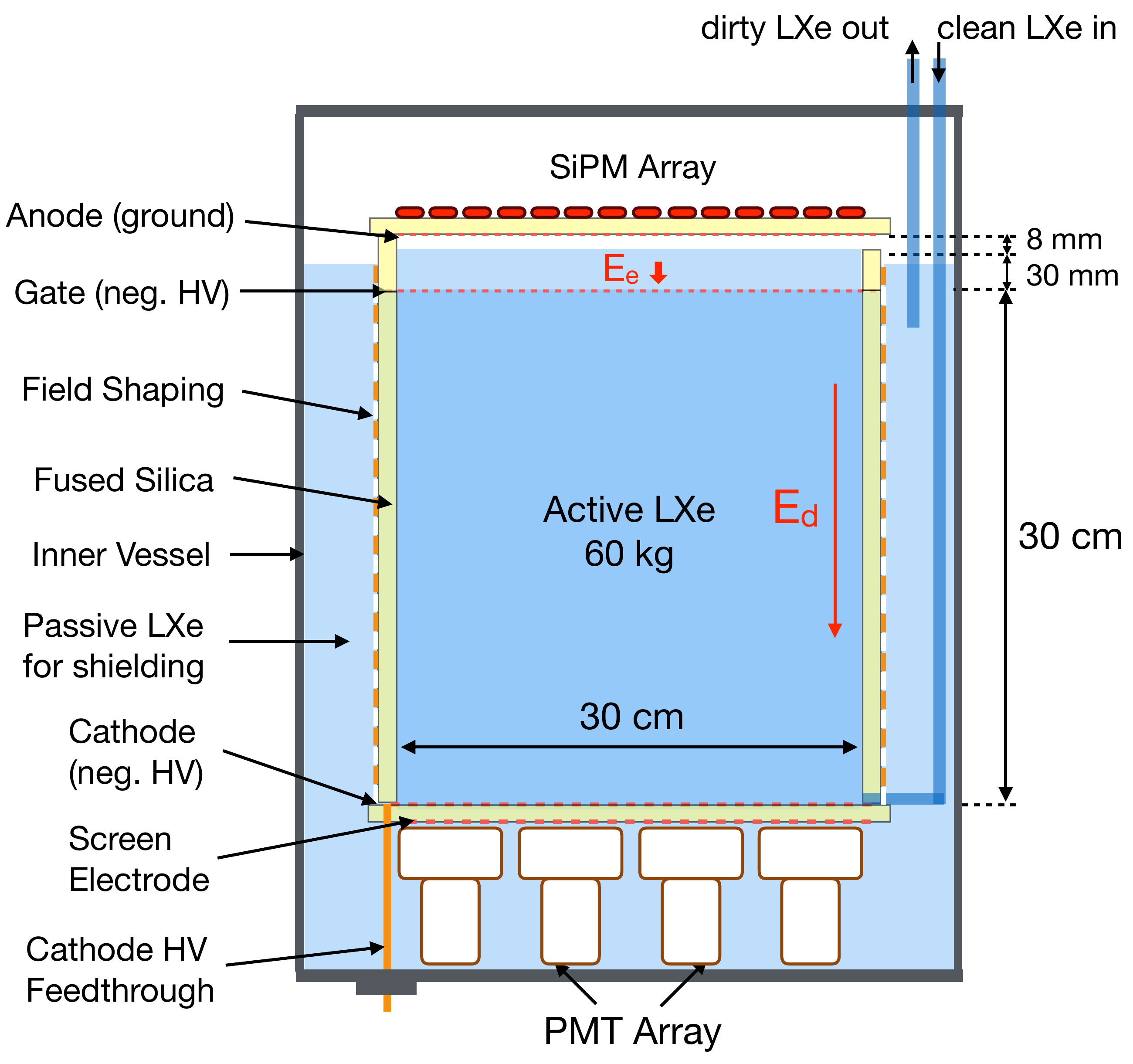}
\end{center}
\caption{(Left) Single and a few electron background rates observed in the XENON10, XENON100 and XENON1T detectors, compared with a benchmark signal rate from 100 MeV dark matter scattering on electrons via a light mediator with a cross section about $10^{-36}\rm~cm^2$ on the ``freeze-in" line in Fig.~\ref{fig:s2onlyprojections} right. (Right) The conceptual design of the LBECA detector containing an active LXe target sealed inside a fused silica enclosure. }
\label{fig:lbecadet}
\end{figure}

\begin{figure*}[!htb]
\begin{center}
\vspace{-3mm}
\begin{tabular}{cc}
\includegraphics[width=0.47\columnwidth]{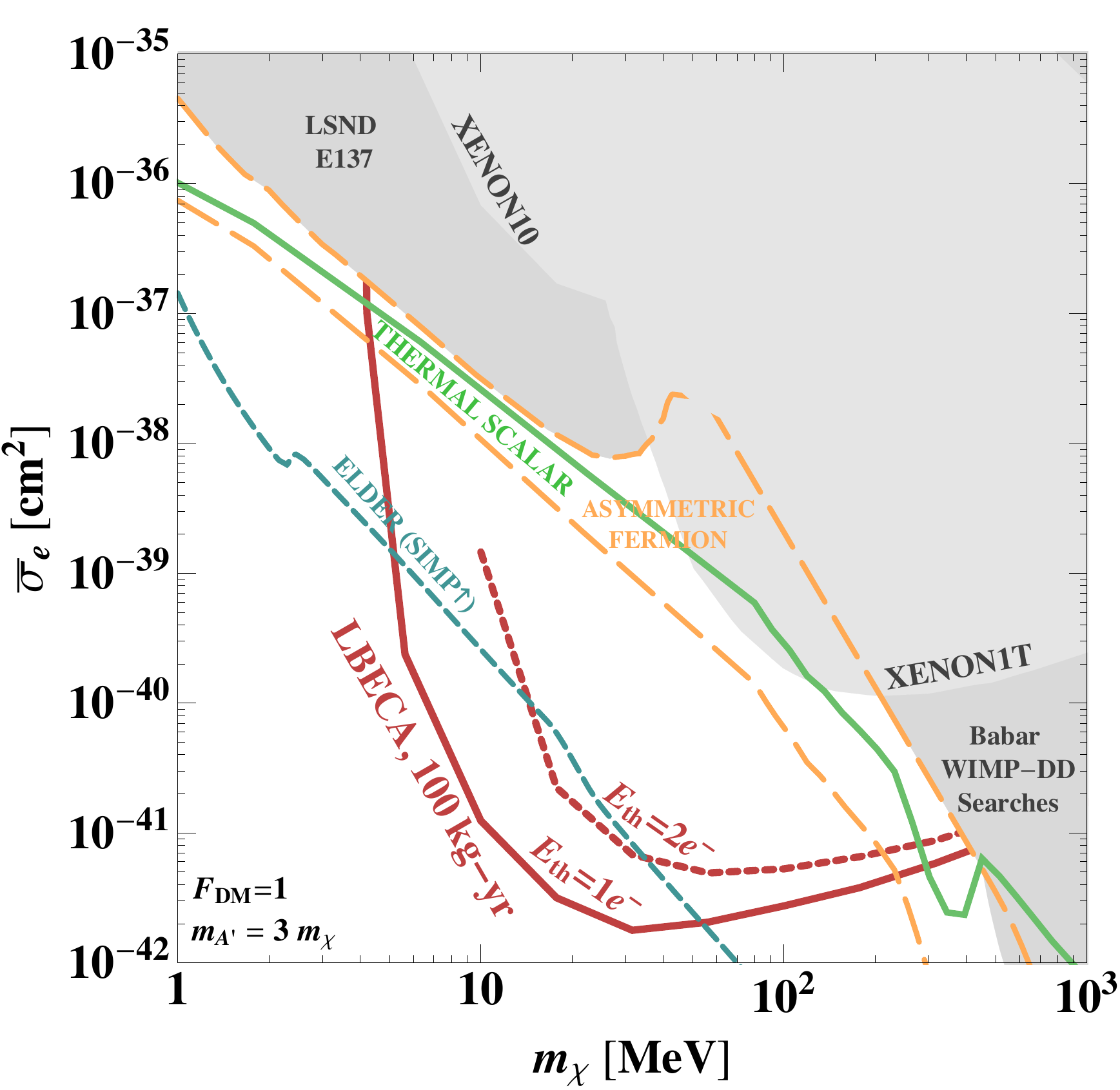} & 
\includegraphics[width=0.47\columnwidth]{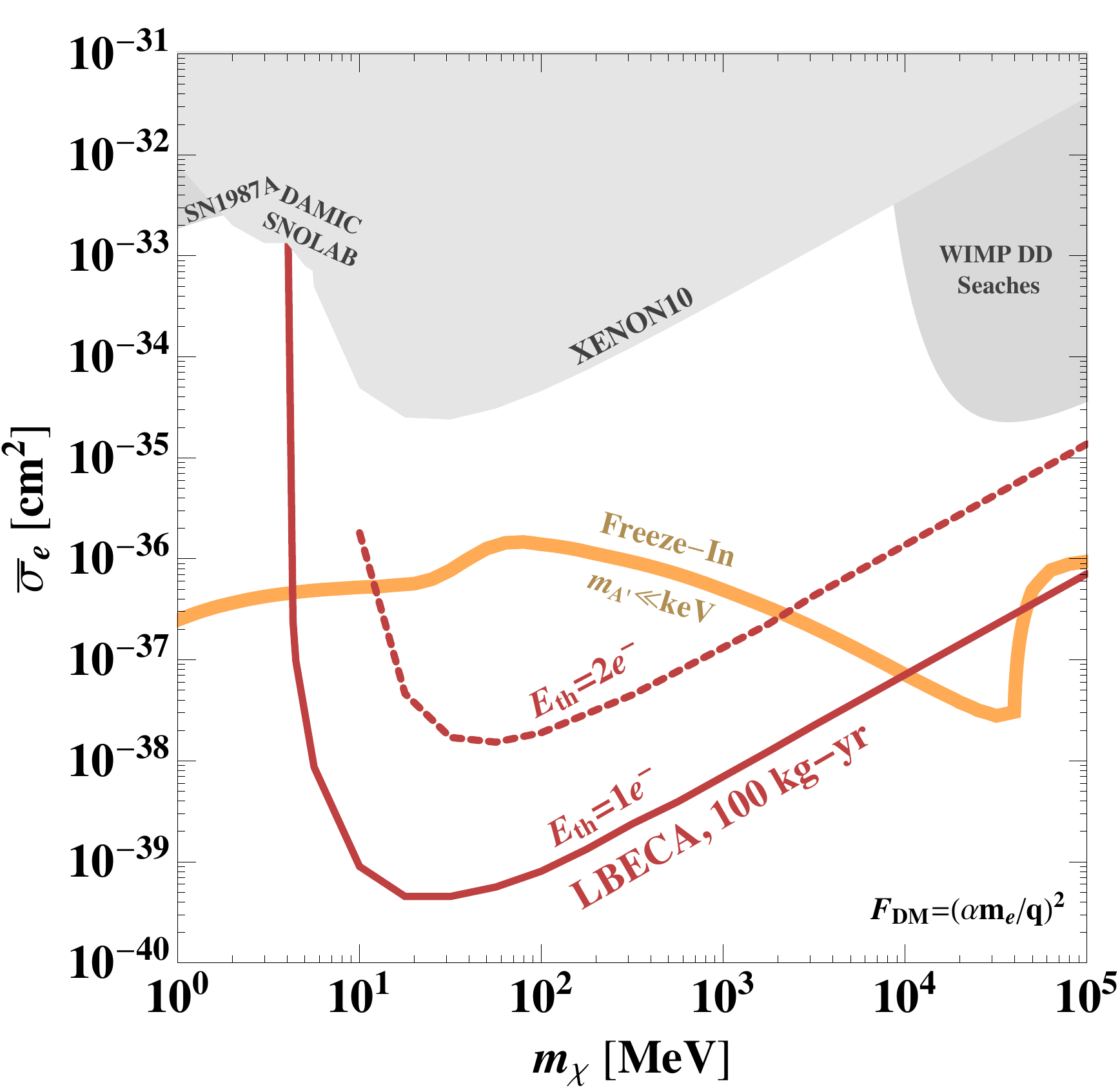}
\end{tabular}
\end{center}
\caption{Red solid (dashed) line shows the sensitivity projection for the LBECA experiment assuming a 100~kg-year exposure and a single-electron (two-electron) threshold without spurious few-electron backgrounds, but including the background from solar neutrinos scattering coherently off nuclei~\cite{Essig:2018tss}.  \textbf{Left:} 
Light-gray regions are constrained from direct-detection experiments~\cite{Essig:2012yx,Essig:2017kqs,Angle:2011th,Aprile:2016wwo,Agnes:2018oej,Tiffenberg:2017aac,Crisler:2018gci,sensei2019,Agnese:2018col}. For the example of a hidden-sector dark matter particle, $\chi$ coupled to a dark photon, $A'$, with $m_{A'}= 3 m_\chi$, we show a darker gray region that combines constraints from accelerator-based searches (LSND, E137, BaBar) and traditional nuclear recoil searches~\cite{Batell:2009di,Essig:2013vha,Batell:2014mga,Essig:2015cda,Essig:2017kqs}.  We also show several benchmark theory targets for which the dark matter can obtain the observed relic abundance~\cite{Hochberg:2014dra,Kuflik:2015isi,Boehm:2003hm,Lin:2011gj,Essig:2015cda}. 
\textbf{Right:} Here, the scattering is mediated by an ultralight particle. Gray regions show constraints from WIMP searches and supernova 1987A in the case of a dark photon mediator~\cite{Essig:2011nj,Chang:2018rso,Essig:2015cda} and the freeze-in benchmark target from~\cite{Essig:2011nj,Chu:2011be,Essig:2015cda,Dvorkin:2019zdi}. 
} 
\label{fig:s2onlyprojections}
\end{figure*}

This proportional scintillation signal gives LXe TPCs the striking sensitivity to any process that can knock out even just a single electron from its shell. While this unique capability is crucial for several exciting physics analyses, these channels are limited by instrumental backgrounds. This problem can be seen in Fig.~\ref{fig:lbecadet} (left): the observed rates from XENON10, XENON100 and XENON1T at about 4 e- are similar, but at a rate about 2 to 3 orders of magnitude higher than that from 100 MeV dark matter scattering on electrons via a light mediator with a cross section about $10^{-36}\rm~cm^2$ on the ``freeze-in" line in Fig.~\ref{fig:s2onlyprojections} (right). The high single-and-a-few electron background rates in these low background liquid xenon detectors provide a challenge but also opportunity to build a dedicated 100-kg scale two-phase xenon detector to significantly suppress these background for a sensitive light dark matter detection.  

One significant source of background comes from photoionization in the bulk liquid xenon or metal surfaces contacting the target. A xenon scintillation photon has 7~eV of energy and can readily photo-ionize exposed metal surfaces or negatively charged impurities in the LXe due to electron attachment, as observed in the XENON100 experiment~\cite{Aprile:2013blg}. Another different electron background is observed in both LUX and XENON1T. The intrinsic time scale of that background is two orders of magnitude longer than that of photoionization, possibly due to electrons trapped at the liquid/gas interface from incomplete extraction~\cite{Sorensen:2017ymt} or long-lived states from residual impurities in the LXe~\cite{Sorensen:2017kpl}. Significant improvement of liquid xenon purity and enhancement of extraction field can reduce these two sources of background. While these improvements are very challenging in multi-ton scale detectors, they can be achieved in a 100-kg scale liquid xenon with improved design.

\section{LBECA Design}
LBECA will be a dual-phase LXe detector optimized for the detection of single- to few-ionization electrons. The detector design is based on the demonstrated LXe TPC technology that has achieved the lowest radioactive background of all direct-search dark matter experiments and is additionally optimized for low ionization-like background rates at energies below the threshold of conventional LXe TPCs. 

A conceptual design of the LBECA detector is shown in Figure~\ref{fig:lbecadet} (right). The active LXe target with total $\sim$60-kg mass is enclosed inside a 30-cm (height) x 30-cm (diameter) fused silica vessel separating the ultra-clean LXe target from external components instrumenting the detector. Purified LXe is fed directly into the bottom of the active target. The LXe overflows out at the top, which defines the liquid-gas xenon interface. Another pipe takes the xenon out to the xenon gas circulation- and purification-system. The clean LXe flowing into the active target will have minimal contamination from the outgassing of other detector components. The passive LXe is only used for shielding and its purity is not important. Such a xenon--circulation loop design ensures the purification efficiency of the liquid xenon in the active target, as demonstrated in our prototype~\cite{lbeca-taup:2019}, as well as in ~\cite{Sato:2019qpr}. Special considerations, such as a very strong extraction field~\cite{LLNL2019_EEE} and infrared light enhanced electron emission, are implemented in the design to ensure complete electron extraction at the liquid-gas interface. The anode and cathode are made from highly transparent conductive mesh, coated on the fused silica windows. High work function material and surface treatment will be done on these electrodes to suppress electron emission. An array of silicon photomultiplier tubes (SiPM) will be installed on the top, viewing the S2 light produced in the gas gap. The high granularity of the SiPM array allows better XY position resolution, compared to that using larger area PMTs. 

\section{LBECA Science Goals and Sensitivity}
LBECA will be sensitive to electronic recoils from dark matter-electron scattering down to dark matter masses of about 5~MeV, which we illustrate in Fig.~\ref{fig:s2onlyprojections}. This sensitivity allows LBECA to probe five specific theory targets with correct predictions of relic-abundance: thermal scalar, asymmetric fermion, SIMP, ELDER and “freeze-in”, which are in a parameter space known as ``The Key Milestone". Moreover, LBECA will be sensitive to bosonic dark matter that is absorbed by electrons for dark matter masses as low as about 9~eV (the atomic ionization energy of xenon)~\cite{An:2014twa,Bloch:2016sjj,Hochberg:2016sqx}. In addition to probing sub-GeV dark matter interactions with electrons, LBECA's unprecedented low-background and ability to detect electrons will also probe sub-GeV dark matter interactions with \textit{nuclei} through the ``Migdal'' effect~\cite{Ibe:2017yqa,Essig:2019xkx,Baxter:2019pnz} and be able to make a measurement of solar $^8$B neutrinos through coherent elastic neutrino-nucleus scattering~\cite{Essig:2018tss}.

This material is based upon work supported by the U.S. Department of Energy, Office of Science, Office of High Energy Physics under Award Number \textbf{DE-SC0018952}.

\section*{References}
\bibliographystyle{hunsrt}
\bibliography{lbeca.bib}

\begin{thebibliography}{10}

\bibitem{Battaglieri:2017aum}
Marco Battaglieri et~al.
\newblock {US Cosmic Visions: New Ideas in Dark Matter 2017: Community Report}.
\newblock 2017, 1707.04591.

\bibitem{BRNreport}
{Department of Energy}.
\newblock {Basic Research Needs for Dark Matter Small Projects New
  Initiatives}.
\newblock 2019.

\bibitem{Essig:2011nj}
Rouven Essig et~al.
\newblock {Direct Detection of Sub-GeV Dark Matter}.
\newblock {\em Phys. Rev.}, D85:076007, 2012, 1108.5383.

\bibitem{Essig:2012yx}
Rouven Essig et~al.
\newblock {First Direct Detection Limits on sub-GeV Dark Matter from XENON10}.
\newblock {\em Phys. Rev. Lett.}, 109:021301, 2012, 1206.2644.

\bibitem{Angle:2011th}
J.~Angle et~al.
\newblock {A search for light dark matter in XENON10 data}.
\newblock {\em Phys. Rev. Lett.}, 107:051301, 2011, 1104.3088.
\newblock [Erratum: Phys. Rev. Lett.110,249901(2013)].

\bibitem{Essig:2017kqs}
Rouven Essig, Tomer Volansky, and Tien-Tien Yu.
\newblock {New Constraints and Prospects for sub-GeV Dark Matter Scattering off
  Electrons in Xenon}.
\newblock {\em Phys. Rev.}, D96(4):043017, 2017, 1703.00910.

\bibitem{Aprile:2016wwo}
E.~Aprile et~al.
\newblock {Low-mass dark matter search using ionization signals in XENON100}.
\newblock {\em Phys. Rev.}, D94(9):092001, 2016, 1605.06262.
\newblock [Erratum: Phys. Rev.D95,no.5,059901(2017)].

\bibitem{Agnes:2018oej}
P.~Agnes et~al.
\newblock {Constraints on Sub-GeV Dark-Matter–Electron Scattering from the
  DarkSide-50 Experiment}.
\newblock {\em Phys. Rev. Lett.}, 121(11):111303, 2018, 1802.06998.

\bibitem{Crisler:2018gci}
Michael Crisler et~al.
\newblock {SENSEI: First Direct-Detection Constraints on sub-GeV Dark Matter
  from a Surface Run}.
\newblock {\em Phys. Rev. Lett.}, 121(6):061803, 2018, 1804.00088.

\bibitem{sensei2019}
Orr Abramoff et~al.
\newblock {SENSEI: Direct-Detection Constraints on Sub-GeV Dark Matter from a
  Shallow Underground Run Using a Prototype Skipper-CCD}.
\newblock {\em Phys. Rev. Lett.}, 122(16):161801, 2019, 1901.10478.

\bibitem{Aguilar-Arevalo:2019wdi}
A.~Aguilar-Arevalo et~al.
\newblock {Constraints on Light Dark Matter Particles Interacting with
  Electrons from DAMIC at SNOLAB}.
\newblock 2019, 1907.12628.

\bibitem{Agnese:2018col}
R.~Agnese et~al.
\newblock {First Dark Matter Constraints from a SuperCDMS Single-Charge
  Sensitive Detector}.
\newblock {\em Phys. Rev. Lett.}, 121(5):051301, 2018, 1804.10697.
\newblock [erratum: Phys. Rev. Lett.122,no.6,069901(2019)].

\bibitem{Aprile:2019xxb}
E.~Aprile et~al.
\newblock {Light Dark Matter Search with Ionization Signals in XENON1T}.
\newblock 2019, 1907.11485.

\bibitem{Aprile:2013blg}
E.~Aprile et~al.
\newblock {Observation and applications of single-electron charge signals in
  the XENON100 experiment}.
\newblock {\em J. Phys.}, G41:035201, 2014, 1311.1088.

\bibitem{Sorensen:2017ymt}
Peter Sorensen.
\newblock {Electron train backgrounds in liquid xenon dark matter search
  detectors are indeed due to thermalization and trapping}.
\newblock 2017, 1702.04805.

\bibitem{Sorensen:2017kpl}
P.~Sorensen and K.~Kamdin.
\newblock {Two distinct components of the delayed single electron noise in
  liquid xenon emission detectors}.
\newblock {\em JINST}, 13(02):P02032, 2018, 1711.07025.

\bibitem{lbeca-taup:2019}
K.~Ni et~al.
\newblock {LBECA: A Low Background Electron Counting Apparatus for Sub-GeV Dark
  Matter Detection}.
\newblock {\em {TAUP 2019,
  http://www-kam2.icrr.u-tokyo.ac.jp/indico/event/3/session/45/contribution/390}}.

\bibitem{Sato:2019qpr}
Kazufumi Sato et~al.
\newblock {Development of Dual-phase Liquid Xenon TPC with a Hermetic Quartz
  Chamber}.
\newblock 2019, 1910.13831.

\bibitem{LLNL2019_EEE}
J.~Xu, S.~Pereverzev, et~al.
\newblock Electron extraction efficiency study for dual-phase xenon dark matter
  experiments.
\newblock {\em Phys. Rev. D}, 99:103024, May 2019.

\bibitem{Essig:2018tss}
Rouven Essig et~al.
\newblock {Solar Neutrinos as a Signal and Background in Direct-Detection
  Experiments Searching for Sub-GeV Dark Matter With Electron Recoils}.
\newblock {\em Phys. Rev.}, D97(9):095029, 2018, 1801.10159.

\bibitem{Tiffenberg:2017aac}
Javier Tiffenberg et~al.
\newblock {Single-electron and single-photon sensitivity with a silicon Skipper
  CCD}.
\newblock {\em Phys. Rev. Lett.}, 119(13):131802, 2017, 1706.00028.

\bibitem{Batell:2009di}
Brian Batell, Maxim Pospelov, and Adam Ritz.
\newblock {Exploring Portals to a Hidden Sector Through Fixed Targets}.
\newblock {\em Phys. Rev.}, D80:095024, 2009, 0906.5614.

\bibitem{Essig:2013vha}
Rouven Essig, Jeremy Mardon, Michele Papucci, Tomer Volansky, and Yi-Ming
  Zhong.
\newblock {Constraining Light Dark Matter with Low-Energy $e^+e^-$ Colliders}.
\newblock {\em JHEP}, 11:167, 2013, 1309.5084.

\bibitem{Batell:2014mga}
Brian Batell, Rouven Essig, and Ze'ev Surujon.
\newblock {Strong Constraints on Sub-GeV Dark Sectors from SLAC Beam Dump
  E137}.
\newblock {\em Phys. Rev. Lett.}, 113(17):171802, 2014, 1406.2698.

\bibitem{Essig:2015cda}
Rouven Essig, Marivi Fernandez-Serra, Jeremy Mardon, Adrian Soto, Tomer
  Volansky, and Tien-Tien Yu.
\newblock {Direct Detection of sub-GeV Dark Matter with Semiconductor Targets}.
\newblock {\em JHEP}, 05:046, 2016, 1509.01598.

\bibitem{Hochberg:2014dra}
Yonit Hochberg, Eric Kuflik, Tomer Volansky, and Jay~G. Wacker.
\newblock {Mechanism for Thermal Relic Dark Matter of Strongly Interacting
  Massive Particles}.
\newblock {\em Phys. Rev. Lett.}, 113:171301, 2014, 1402.5143.

\bibitem{Kuflik:2015isi}
Eric Kuflik, Maxim Perelstein, Nicolas Rey-Le Lorier, and Yu-Dai Tsai.
\newblock {Elastically Decoupling Dark Matter}.
\newblock {\em Phys. Rev. Lett.}, 116(22):221302, 2016, 1512.04545.

\bibitem{Boehm:2003hm}
C.~Boehm and Pierre Fayet.
\newblock {Scalar dark matter candidates}.
\newblock {\em Nucl. Phys.}, B683:219--263, 2004, hep-ph/0305261.

\bibitem{Lin:2011gj}
Tongyan Lin, Hai-Bo Yu, and Kathryn~M. Zurek.
\newblock {On Symmetric and Asymmetric Light Dark Matter}.
\newblock {\em Phys. Rev.}, D85:063503, 2012, 1111.0293.

\bibitem{Chang:2018rso}
Jae~Hyeok Chang, Rouven Essig, and Samuel~D. McDermott.
\newblock {Supernova 1987A Constraints on Sub-GeV Dark Sectors, Millicharged
  Particles, the QCD Axion, and an Axion-like Particle}.
\newblock {\em JHEP}, 09:051, 2018, 1803.00993.

\bibitem{Chu:2011be}
Xiaoyong Chu, Thomas Hambye, and Michel H.~G. Tytgat.
\newblock {The Four Basic Ways of Creating Dark Matter Through a Portal}.
\newblock {\em JCAP}, 1205:034, 2012, 1112.0493.

\bibitem{Dvorkin:2019zdi}
Cora Dvorkin, Tongyan Lin, and Katelin Schutz.
\newblock {Making dark matter out of light: freeze-in from plasma effects}.
\newblock 2019, 1902.08623.

\bibitem{An:2014twa}
Haipeng An, Maxim Pospelov, Josef Pradler, and Adam Ritz.
\newblock {Direct Detection Constraints on Dark Photon Dark Matter}.
\newblock {\em Phys. Lett.}, B747:331--338, 2015, 1412.8378.

\bibitem{Bloch:2016sjj}
Itay~M. Bloch, Rouven Essig, Kohsaku Tobioka, Tomer Volansky, and Tien-Tien Yu.
\newblock {Searching for Dark Absorption with Direct Detection Experiments}.
\newblock {\em JHEP}, 06:087, 2017, 1608.02123.

\bibitem{Hochberg:2016sqx}
Yonit Hochberg, Tongyan Lin, and Kathryn~M. Zurek.
\newblock {Absorption of light dark matter in semiconductors}.
\newblock {\em Phys. Rev.}, D95(2):023013, 2017, 1608.01994.

\bibitem{Ibe:2017yqa}
Masahiro Ibe, Wakutaka Nakano, Yutaro Shoji, and Kazumine Suzuki.
\newblock {Migdal Effect in Dark Matter Direct Detection Experiments}.
\newblock {\em JHEP}, 03:194, 2018, 1707.07258.

\bibitem{Essig:2019xkx}
Rouven Essig, Josef Pradler, Mukul Sholapurkar, and Tien-Tien Yu.
\newblock {On the relation between Migdal effect and dark matter-electron
  scattering in atoms and semiconductors}.
\newblock 2019, 1908.10881.

\bibitem{Baxter:2019pnz}
Daniel Baxter, Yonatan Kahn, and Gordan Krnjaic.
\newblock {Electron Ionization via Dark Matter-Electron Scattering and the
  Migdal Effect}.
\newblock 2019, 1908.00012.

\end{thebibliography}

\end{document}